\documentclass[12pt,a4]{article}
\usepackage{amsmath,epsfig}
\begin{document}
\begin{titlepage}
  \begin{flushright}
    Wisc-Ex-99-352\\
    H. Hu, J. Nielsen\\
    June 1, 1999\\
  \end{flushright}
  \vspace{20mm}
  \begin{center}
    {\bf \Large Analytic Confidence Level Calculations using \\*[0.3cm]
      the Likelihood Ratio and Fourier Transform} \vskip 1.6 cm
    {\sc Hongbo Hu and Jason Nielsen}\footnote[1]{electronic mail:
    hu@wisconsin.cern.ch, nielsen@wisconsin.cern.ch} \\  
    {\it University of Wisconsin-Madison, Wisconsin, USA}
  \end{center}

  \vspace{10mm}

  \begin{abstract} 
The interpretation of new particle search results
involves a confidence level calculation on either the
discovery hypothesis or the background-only (``null'')
hypothesis.  A typical approach uses toy Monte Carlo
experiments to build an expected experiment estimator
distribution against which an observed experiment's estimator
may be compared.  In this note, a new approach is presented
which calculates analytically the experiment estimator
distribution via a Fourier transform, using the likelihood
ratio as an ordering estimator.  The analytic approach enjoys
an enormous speed advantage over the toy Monte Carlo method,
making it possible to quickly and precisely calculate
confidence level results.
  \end{abstract}
\end{titlepage}

\section{Introduction}
\label{sect:intro}
A consistently recurring topic at LEP2 has been the interpretation and
combination of results from searches for new particles.  The
fundamental task is to interpret the collected dataset in the context
of two complementary hypotheses.  The first hypothesis -- the {\it
  null hypothesis} -- is that the dataset is compatible with
non-signal Standard Model background production alone, and the second
is that the dataset is compatible with the sum of signal and Standard
Model background production.  In most cases, the search for new
particles proceeds via several parallel searches for final states.
The results from all of these subchannels are then combined to produce
a final result.

All existing confidence level calculations follow the same general
strategy \cite{Janot, Junk, ARead}.  A test statistic or {\it
  estimator} is constructed to quantify the ``signal-ness'' of a real
or simulated experiment.  The ``signal-ness'' of a single observed
experiment leads to the confidence level on, for example, the null
hypothesis that the observed experiment is incompatible with signal
and background both being produced.  Most calculation methods use an
ensemble of toy Monte Carlo experiments to generate the estimator
distribution against which the observed experiment is compared.  This
generation can be rather time-consuming when the number of toy Monte Carlo
experiments is great (as it must be for high precision calculations)
or if the number of signal and background expected for each experiment
is great (as it is for the case of searches optimized to use
background subtraction).

In this note, we present an improved method for calculating confidence
levels in the context of searches for new particles.  Specifically,
when the likelihood ratio is used as an estimator, the experiment
estimator distribution may be calculated analytically with the Fourier
transform.  With this approach, the disadvantage of toy Monte Carlo
experiments is avoided.  The analytic method offers several advantages
over existing methods, the most dramatic of which is the increase in
calculation speed and precision.

\section{Likelihood ratio estimator for searches}

The likelihood ratio estimator is the ratio of the probabilities of
observing an event under two search hypotheses.  The
estimator for a single experiment is
\begin{equation}
  E = C \frac{{\mathcal L}_{s+b}}{{\mathcal L}_{b}}.
\end{equation}

Here ${\mathcal L}_{s+b}$ is the probability density function for
signal+background experiments and ${\mathcal L}_{b}$ is the
probability density function for background-only experiments.  Because
the constant factor $C$ appears in each event's estimator, it does not
affect the ordering of the estimators -- an event cannot become more
signal-like by choosing a different $C$.  For clarity in this note,
the constant is chosen to be $e^s$, where $s$ is the expected number
of signal events.~\footnote{When considering the two production
  hypotheses and calculating an exclusion, the expected signal $s$ is
  uniquely determined by the cross section.  If the cross section is
  not fixed, then $e^s$ is not constant, and $C$ may be set to unity.}
For the simplest case of event counting with no discriminant variables
(or, equivalently, with perfectly non-discriminating variables), the
estimator can be calculated with Poisson probabilities alone.  In
practice, not every event is equally signal-like.  Each search may
have one or more event variables that discriminate between signal-like
and background-like events.  For the general case, the probabilities
${\mathcal L}_{s+b}$ and ${\mathcal L}_{b}$ are functions of the
observed events' measured variables.

As an example, consider a search using one discriminant variable $m$,
the reconstructed Higgs mass.  The signal and background have
different probability density functions of $m$, defined as $f_{s}(m)$
and $f_{b}(m)$, respectively.  (For searches with more than one
discriminant variable, $m$ is replaced by a vector of discriminant
variables $\overrightarrow{x}$.)  It is then straightforward to
calculate ${\mathcal L}_{s+b}$ and ${\mathcal L}_{b}$ for a single
event, taking into account the event weighting coming from the
discriminant variables:
\begin{equation}
  E = e^{s} \frac{P_{s+b}}{P_{b}} =
  e^{s} \frac{e^{-(s+b)} \left[s f_{s}(m) + b f_{b}(m)\right]}{e^{-b}
    \left[b f_{b}(m)\right]}.
\end{equation}

The likelihood ratio estimator can be shown to maximize the discovery
potential and exclusion potential of a search for new particles
\cite{ARead}.  Such an estimator, both with and without discriminant
variables, has been used successfully by the LEP2 collaborations to
calculate confidence levels for searches \cite{Junk, ARead}.

\section{Ensemble estimator distributions via Fast Fourier Transform (FFT)}
One way to form an estimator for an ensemble of events is to generate
a large number of toy Monte Carlo experiments, each experiment having
a number of events generated from a Poisson distribution.  Another way
is to analytically compute the probability density function of the
ensemble estimator given the probability density function of the event
estimator.  The discussion of this section pursues the latter
approach.

The likelihood ratio estimator is a multiplicative estimator.  This
means the estimator for an ensemble of events is formed by multiplying
the individual event estimators.  Alternatively, the
logarithms of the estimators may be summed.  In the following
derivation, $F = \ln E$, where $E$ is the likelihood ratio estimator.

For an experiment with 0 events observed, the estimator is trivial: 
\begin{eqnarray}
  E & = & e^{s} \frac{e^{-(s+b)}}{e^{-b}} = 1 \\
  F & = & 0 \\
  \rho_{0}(F) & = & \delta(F),
\end{eqnarray}
where $\rho_0 (F)$ is the probability density function of $F$ for
experiments with 0 observed events.

For an experiment with exactly one event, the estimator is, again
using the reconstructed Higgs mass $m$, 
\begin{eqnarray}
  E & = & e^{s} \frac{e^{-(s+b)} \left[s f_{s}(m) + b f_{b}(m)\right]}{e^{-b}
    \left[b f_{b}(m)\right]}, \\
  F & = & \ln \frac{s f_{s}(m) + b f_{b}(m)}{b f_{b}(m)},
\end{eqnarray}
and the probability density function of $F$ is defined as $\rho_1 (F)$.

For an experiment with exactly two events, the estimators of the two
events are multiplied to form an event estimator.  If the
reconstructed Higgs masses of the two events are $m_1$ and $m_2$, then 
\begin{eqnarray}
  E & = & \frac{\left[s f_{s}(m_1) + b f_{b}(m_1)\right] \left[s f_{s}(m_2) + b
      f_{b}(m_2)\right]}{\left[b f_{b}(m_1)\right]\left[b
      f_{b}(m_2)\right]} \\
  F & = & \ln \frac{s f_{s}(m_1) + b f_{b}(m_1)}{b f_{b}(m_1)} + 
  \ln \frac{s f_{s}(m_2) + b f_{b}(m_2)}{b f_{b}(m_2)}.
\end{eqnarray}
The probability density function for exactly two particles $\rho_2(F)$
is simply the convolution of $\rho_1 (F)$ with itself: 
\begin{eqnarray}
  \rho_2(F) & = & \iint \rho_1(F_1) \rho_1(F_2)
  \delta(F - F_1 - F_2) dF_1 dF_2 \\
  & = & \rho_1(F) \otimes \rho_1(F).  
\end{eqnarray}

The generalization to the case of $n$ events is straightforward and
encouraging:
\begin{eqnarray}
  E & = & \prod^{n}_{i=1} \frac{s f_{s}(m_i) + b f_{b}(m_i)}{b f_{b}(m_i)} \\
  F & = & \sum^{n}_{i=1} \ln \frac{s f_{s}(m_i) + b f_{b}(m_i)}{b
  f_{b}(m_i)} \\
  \rho_n (F) & = & \idotsint \prod^{n}_{i=1} \left[ \rho_1(F_i) dF_i
  \right] \delta \left( F - \sum^{n}_{i=1} F_i \right) \\
  & = & \underbrace {\rho_1 (F) \otimes \dots \otimes \rho_1 (F)}_{n\ 
    \text{times}}.
\end{eqnarray}

Next, the convolution of $\rho_1 (F)$ is rendered manageable by an
application of the relationship between the convolution and the
Fourier transform.

If $A(F) = B(F) \otimes C(F)$, then the Fourier transforms of $A$,
$B$, and $C$ satisfy
\begin{equation}
  \overline{A(G)} = \overline{B(G)} \cdot \overline{C(G)}.
\end{equation}
This allows the convolution to be expressed as a simple power:
\begin{equation}
  \overline{\rho_n(G)} = \left[ \overline{\rho_1(G)} \right]^n.
\end{equation}
Note this equation holds even for $n=0$, since $\overline{\rho_0(G)} = 1$.
For any practical computation, the analytic Fourier transform can be
approximated by a numerical Fast Fourier Transform (FFT) \cite{FFT}.

How does this help to determine $\rho_{s+b}$ and $\rho_{b}$?  The
probability density function for an experiment estimator with $s$ expected
signal and $b$ expected background events is
\begin{equation}
  \rho_{s+b}(F) = \sum^{\infty}_{n=0} e^{-(s+b)} \frac{(s+b)^n}{n!}
  \rho_n(F),
\end{equation}
where $n$ is the number of events observed in the experiment.
Upon Fourier transformation, this becomes
\begin{eqnarray}
  \overline{\rho_{s+b}(G)} & = & \sum^{\infty}_{n=0} e^{-(s+b)} \frac{(s+b)^n}{n!}
  \overline{\rho_n(G)} \\
  & = & \sum^{\infty}_{n=0} e^{-(s+b)} \frac{(s+b)^n}{n!} \left
    [ \overline{\rho_1(G)} \right]^n
\end{eqnarray}

\begin{equation}
  \boxed{\overline{\rho_{s+b}(G)} = e^{(s+b)\left[
 \overline{\rho_1(G)} - 1 \right]}}
\end{equation}
The function $\rho_{s+b}(F)$ may then be recovered by using the inverse
transform.  In general, this relation holds for any
multiplicative estimator.

This final relation means that the probability density function for an
arbitrary number of expected signal and background events can be
calculated analytically once the probability density function of the
estimator is known for a single event.  This calculation is therefore
just as fast for high background searches as for low background
searches.  In particular, it holds great promise for Higgs searches
which, due to use of background subtraction and discriminant variables,
are optimized to higher background levels than they have been in the past.

Two examples will provide practical proof of the principle.  For the
first, assume a
hypothetical estimator results in a probability density function of
simple Gaussian form 
\begin{equation}
  \rho_1 (F) = \frac{1}{\sqrt{2 \pi \sigma}}e^{- \frac{(x-\mu)^2}{2
  \sigma^2}},
\end{equation} 
where $\sigma = 0.2$ and $\mu = 2.0$.  For an expected $s+b=20.0$,
both the FFT method and the toy Monte Carlo method are used to evolute the
event estimator probability density function to an experiment
estimator probability density function.  The agreement between the two
methods (Fig.~1) is striking.
\begin{figure}[htbp]
  \begin{center}
    \epsfig{file=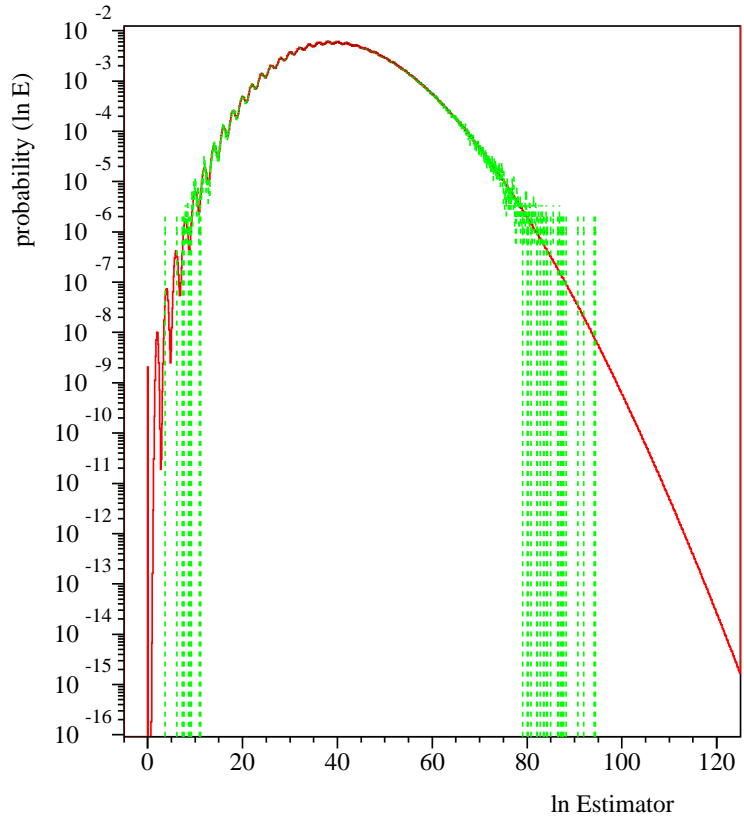,width=0.92\textwidth}
    \caption{The experiment estimator probability density function
      for a simple event estimator probability function calculated
      with the FFT method (solid red line) and the toy Monte Carlo
      method (dashed green line).  Error bars associated with the Monte
      Carlo method are due to limited statistics.}
  \end{center}
  \label{simple}
\end{figure}
The higher precision of the FFT method is apparent, even when compared
to 1 million toy Monte Carlo experiments.  The periodic structure is
due to the discontinuous Poisson distribution being convolved with a
narrow event estimator probability function.  In particular, the peak
at $\ln E = 0$ corresponds to the probability that exactly zero events
be observed ($e^{-(s+b)} = 2.1 \times 10^{-9}$).  The precision of the
toy Monte Carlo method is limited by the number of Monte Carlo
experiments, while the precision of the FFT method is limited only by
computer precision.

For the second example, a more realistic estimator is calculated using
a discriminant variable distribution from an imaginary $\text{HZ}
\rightarrow \text{H} \tau \tau$ search.  The variable used here is the
reconstructed Higgs mass of the event.  This estimator's probability
density function is then calculated for an experiment with $s=5$ and
$b=3$ expected events (Fig.~2).
\begin{figure}[htbp]
  \begin{center}
    \epsfig{file=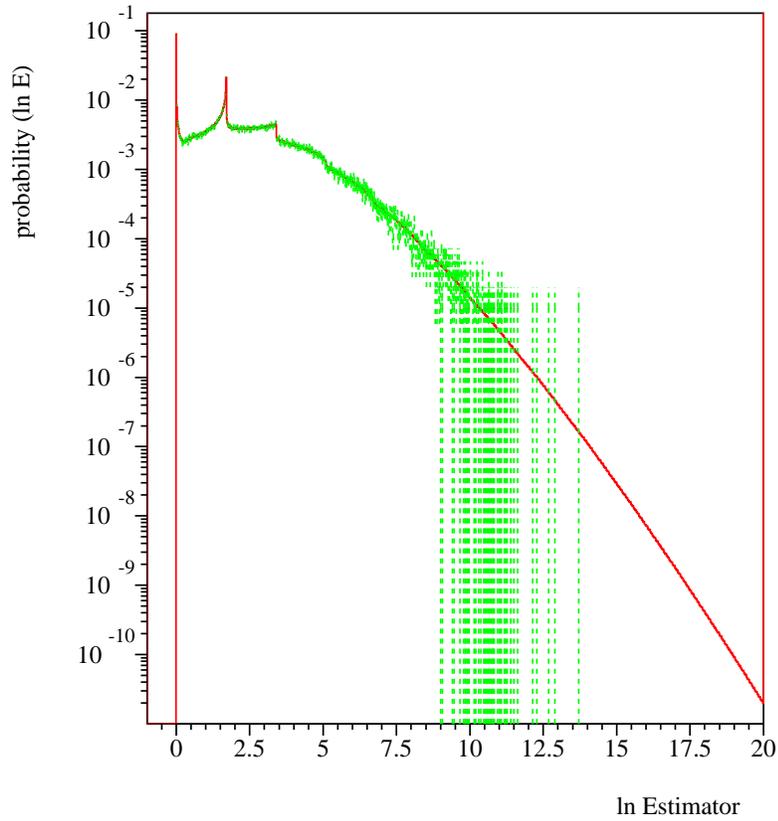,width=0.92\textwidth}
    \caption{The experiment estimator probability density function for
      an estimator based on reconstructed Higgs mass in $\text{HZ}
      \rightarrow \text{H} \tau \tau$ searches.  The result from the
      FFT method is the solid red line, and the result from the toy
      Monte Carlo method is the dashed green line.}
  \end{center}
  \label{tt}
\end{figure}
Again, the two methods agree well in regions where
the toy Monte Carlo method is useful.

These examples support the mathematical proof of the FFT method
described above.  Because the final calculations $c_{s+b}$ and $c_b$
are simply integrals of the experiment estimator probability density
function, any confidence levels calculated with the FFT method and the toy
Monte Carlo method are identical.  The examples also show the precision
achievable with the FFT method, a precision that will be important when
testing discovery hypotheses at the $5\sigma = 5 \times 10^{-7}$ level.

\section{Combining results from several searches}

Given the multiplicative properties of the likelihood ratio estimator,
the combination of several search channels proceeds intuitively.  The
estimator for any combination of events is simply the product of the
individual event estimators.  Consequently, construction of the
estimator probability density function for the combination of channels
parallels the construction of the estimator probability density
function for the combination of events in a single channel.  In
particular, for a combination with $N$ search channels:
\begin{eqnarray}
\overline{\rho_{s+b}(G)} & = & \prod^{N}_{j=1} \overline{\rho_{s+b}^j
  (G)} \\
 & = & e^{\sum^{N}_{j=1} (s_j + b_j) \left[ \overline{\rho_1^j (G)} -
  1 \right]}
\end{eqnarray}  

Due to the strictly multiplicative nature of the estimator, this
combination method is internally consistent.  No matter how subsets of
the combinations are rearranged ({\it i.e.}, combining channels in
different orders, combining different subsets of data runs), the
result of the combination does not change.

Once a results are obtained for $\rho_{s+b} (F)$ and $\rho_b (F)$,
simple integration gives the confidence coefficients $c_{s+b}$ and
$c_b$.  From this point, confidence levels for the two search
hypotheses may be calculated in a number of ways
\cite{Junk,Helene,SE}.  Those straightforward calculations are outside
the scope of this note.

\section{Final remarks and conclusions}

A few short remarks conclude this note and emphasize the advantages of
calculations using the likelihood ratio with the Fast Fourier
Transform (FFT) method.

\begin{enumerate}
\item{ The likelihood ratio estimator is an optimal ordering estimator
    for maximizing both discovery and exclusion potential.  Such an
    estimator can only improve the discovery or exclusion potential of
    a search.}
  
\item{ As a multiplicative estimator, the likelihood ratio estimator
    ensures internal consistency when results are combined.  For
    example, if the dataset is split into several smaller pieces, the
    combined result always remains the same.}
  
\item{ The probability density function of an ensemble estimator may
    be calculated analytically from the event estimator probability
    density function.  Avoiding toy Monte Carlo generation brings
    revolutionary advances in speed and precision.  For a $\text{HZ}
    \rightarrow \text{4-jets}$ search with 25 expected background
    events, a full confidence level calculation with $2^{18}$ toy MC
    experiments and 60 Higgs mass hypotheses takes approximately
    fifteen CPU hours.  By contrast, the same calculation using the
    FFT method takes approximately two CPU minutes.  This discrepancy
    only increases as the required confidence level precision and the
    number of toy MC experiments increase.  For example, confidence
    level calculations for discovery at the $5 \sigma$ level would
    require ${\mathcal O}(10^{8})$ toy MC experiments.  Given the
    approximately linear scaling of calculating time with number of
    toy experiments, such a calculation would take up almost a year in
    the 4-jet channel alone!  The precision of the analytic FFT method
    is more than sufficient for a $5 \sigma$ discovery.}
\end{enumerate}

A fast confidence level calculation makes possible studies that might
have otherwise been too CPU-intensive with the toy MC method.  These include
studies of improvements in the event selections, of various working
points, and of systematic errors and their effects, among others.  A
precise calculation makes possible rejection of null hypotheses at the
level necessary for discovery.

The marriage of the likelihood ratio estimator and the FFT method
seems well-suited for producing extremely fast and precise confidence
level results, and the flexibility and ease of use of the {\tt clfft}
package should make this a powerful tool in interpreting searches for
new particles.

\end{document}